\documentclass[aps,preprint]{revtex4}%
\usepackage{amsfonts}
\usepackage{amsmath}
\usepackage{amssymb}
\usepackage{graphicx}%
\begin{document}
\preprint{ }
\title[G\"{o}ppert-Mayer and the SFA]{Low-frequency failure of the G\"{o}ppert-Mayer gauge transformation and
consequences for the Strong-Field Approximation}
\author{H. R. Reiss}
\affiliation{Max Born Institute, 12489 Berlin, Germany}
\affiliation{American University, Washington, DC 20016-8058, USA}


\begin{abstract}
The G\"{o}ppert-Mayer (GM) gauge transformation, of central importance in
atomic, molecular, and optical physics since it connects the length gauge and
the velocity gauge, becomes unphysical as the field frequency declines towards
zero. This is not consequential for theories of transverse fields, but it is
the underlying reason for the failure of gauge invariance in the
dipole-approximation version of the Strong-Field Approximation (SFA). This
failure of the GM gauge transformation explains why the length gauge is
preferred in analytical approximation methods for fields that possess a
constant electric field as a zero-frequency limit.

\end{abstract}
\date[9 September 2015]{}
\maketitle

The Strong-Field Approximation (SFA) is the basic analytical method for the
treatment of the interaction of nonperturbatively strong laser fields with
atoms and molecules, but it is known to be gauge-dependent. It is important to
note from the outset that results obtained herein apply only to theories that
make \textit{a priori} use of the dipole approximation. Such theories are
identifiable by the fact that a zero-frequency limit exists, and that this
limit corresponds to a constant electric field. In other words, this work
applies only to longitudinal fields. Thus theories that are derived from
propagating-wave formalisms are excluded, since such transverse-field theories
have extremely low-frequency radio waves as a low frequency limit.

Gauge dependence of the SFA is probably shown most clearly in Ref. \cite{bmb},
where the length gauge (LG) results are plausible, but the velocity gauge (VG)
results are not. This long-known lack of gauge invariance has led to
statements of alarm, such as \textquotedblleft\textit{... the SFA is not gauge
invariant, which is really bad news for a theory.}.\textquotedblright%
\ \cite{mishajmo} (emphasis from the original.) Another expression of concern
is: \textquotedblleft...how can a noninvariant theory be used for the
calculation of observables?\textquotedblright\ \cite{poprtun}.

The approach taken here for examination of the gauge problem is entirely
general, with no dependence on the particular properties of any problem or
class of problems beyond the statement that the field is treated as a
longitudinal field (or the equivalent statement that the field has a
zero-frequency limit corresponding to a static electric field). We start with
a re-derivation of the known result \cite{hrjmo,hrtun} that the static
electric field can be described only within a unique gauge if all physical
constraints are to be satisfied. A nominally alternative gauge is discarded on
the grounds that it violates the physical condition that a charged particle in
a static electric field represents a system for which total energy is
conserved. It is then shown that this unphysical gauge arises from a
G\"{o}ppert-Mayer (GM) gauge transformation from the length gauge to the
velocity gauge as applied to an oscillatory electric field in the zero
frequency limit. This establishes the unphysical nature of the GM gauge
transformation when zero frequency is a possibility. This is consequential in
that photoelectron spectra that extend to zero frequency are a necessary part
of any strong-field, nonperturbative problem. The GM gauge transformation from
the LG to the VG is thus shown to be unphysical in the zero-frequency limit,
leaving the LG as the only physical alternative. The two constraints of strong
fields and the accessibility of a zero-frequency limit are all that is
necessary to confirm the LG as the only physical gauge for the SFA in the form
appropriate to oscillatory electric fields.

Consider a static electric field with the amplitude $E_{0}$. It is known from
electrostatics that this field can be specified by the scalar and vector
potentials%
\begin{equation}
\phi=-\mathbf{r\cdot E}_{0},\quad\mathbf{A}=0. \label{a}%
\end{equation}
A gauge transformation can be accomplished by a scalar generating function
$\Lambda$ subject only to the constraint that the generating function satisfy
the homogeneous wave equation%
\begin{equation}
\partial^{\mu}\partial_{\mu}\Lambda=0. \label{b}%
\end{equation}
The 4-vector potential following from a gauge transformation is%
\begin{equation}
\widetilde{A}^{\mu}=A^{\mu}+\partial^{\mu}\Lambda, \label{c}%
\end{equation}
which is equivalent to the transformed scalar and 3-vector potentials%
\begin{align}
\widetilde{\phi}  &  =\phi+\frac{1}{c}\partial_{t}\Lambda,\label{d}\\
\widetilde{\mathbf{A}}  &  =\mathbf{A}-\mathbf{\nabla}\Lambda. \label{e}%
\end{align}
It is well-known that the representation of a static electric field by a
scalar potential alone, as in Eq.(\ref{a}) can be gauge-transformed so that
the field can be described by a vector potential alone by using the generating
function%
\begin{equation}
\Lambda=ct\mathbf{r\cdot E}_{0}, \label{f}%
\end{equation}
which leads to the new potentials%
\begin{equation}
\widetilde{\phi}=0,\quad\widetilde{\mathbf{A}}=-ct\mathbf{E}_{0}. \label{g}%
\end{equation}
The potentials in Eq. (\ref{g}) are unphysical in the sense that a charged
particle subject to those potentials is described by Lagrangian and
Hamiltonian functions that possess explicit time dependence; and explicit time
dependence of these system functions is a clear indicator that total energy is
not conserved. This contrasts with the time independence of the potentials in
Eq.(\ref{a}), signifying energy conservation.

The formal foundations for Noether's Theorem connecting symmetries with
physical conservation laws are expressed in terms of the Lagrangian function.
(See, for example, Ref. \cite{goldstein}.) The potentials (\ref{a}) lead to a
Lagrangian that has no explicit dependence on time, and thereby demonstrates
energy conservation, whereas the potentials (\ref{g}) signify a Lagrangian
that depends explicitly on the time, and is thus unphysical.

The GM gauge transformation is usually expressed in terms of the vector
potential that arises after the transformation. That is, the generator of the
GM gauge transformation is usually written as%
\begin{equation}
\Lambda^{GM}=-\mathbf{r\cdot}\widetilde{\mathbf{A}}.\label{g1}%
\end{equation}
This is exactly what follows from Eqs. (\ref{f}) and (\ref{g}), so the above
discussion amounts to concluding that the GM gauge transformation is
unphysical when $\omega=0$.

Problems described by nonperturbative methods such as tunneling methods
\cite{kel,nr,ppt,adk}, have spectra that are always inclusive of zero
frequency. This is straightforward to describe within the LG, but the
extension to $\omega\rightarrow0$ defies treatment within the VG.

The failure of the GM gauge transformation has no significance for transverse
fields, such as laser fields. Such fields are propagating fields that do not
have a zero frequency limit in the same sense as longitudinal fields.
Propagating fields have extremely low-frequency radio fields as the limit when
$\omega\rightarrow0$ \cite{hr101,hrtun}. The limit point of $\omega=0$ cannot
be achieved for a variety of (inter-related) reasons: propagation is not
defined when $\omega=0$; the magnetic field must always have the same
magnitude as the electric field (in Gaussian units), so it can never be set to
zero when the electric field is nonzero; $\omega\rightarrow0$ implies
wavelength $\lambda\rightarrow\infty$; the ponderomotive energy $U_{p}$ for a
transverse field is proportional to $1/\omega^{2}$, so infinite energy must be
supplied; there is no gauge freedom at all for propagating fields
\cite{hrjmo,hrtun}; and so on.

The overall conclusion is that the LG is the sole physical gauge for
oscillatory electric fields when the zero-frequency limit must be considered.

I thank Prof. D. Bauer of Rostock University for useful discussions.

\end{document}